\documentstyle[12pt,a4wide]{article}
\newcommand{\be}{\begin{equation}}
\newcommand{\ee}{\end{equation}}
\newcommand{\bea}{\begin{eqnarray}}
\newcommand{\eea}{\end{eqnarray}}
\newcommand{\bean}{\begin{eqnarray*}}

\newcommand{\eean}{\end{eqnarray*}}
\font\upright=cmu10 scaled\magstep1
\font\sans=cmss10
\newcommand{\ssf}{\sans}
\newcommand{\stroke}{\vrule height8pt width0.4pt depth-0.1pt}
\newcommand{\Z}{\hbox{\upright\rlap{\ssf Z}\kern 2.7pt {\ssf Z}}}

\newcommand{\C}{{\rlap{\rlap{C}\kern 3.8pt\stroke}\phantom{C}}}
\newcommand{\R}{\hbox{\upright\rlap{I}\kern 1.7pt R}}
\newcommand{\CP}{\C{\upright\rlap{I}\kern 1.5pt P}}
\newcommand{\PP}{\hbox{\upright\rlap{I}\kern 1.5pt P}}

\newcommand{\identity}{{\upright\rlap{1}\kern 2.0pt 1}}

\newcommand{\HH}{\mbox{\hbox{\upright\rlap{I}\kern 1.7pt H}}}
\newcommand{\kk}{\kappa}
\newcommand{\tr}{\mbox{tr}}

\newcommand{\fr}{\frac}

\newcommand{\ra}{\rightarrow}
\newcommand{\al}{\alpha}
\newcommand{\bt}{\beta}
\newcommand{\pr}{\partial}
\newcommand{\hs}{\hspace{5mm}}

\newcommand{\dg}{\dagger}

\newcommand{\acc}{\\[3mm]}

\newcommand{\zb}{{\bar z}}
\newcommand{\pp}{\Delta}

\font\mybb=msbm10 at 11pt

\def\bb#1{\hbox{\mybb#1}}

\def\bC {\bb{C}}

\renewcommand{\CP}{\bC {\rm P}}

\input{epsf}

\begin{document}

\begin{center}{\bf
Non-BPS Black Holes in ${\cal N}=4$ Supersymmetric Yang-Mills Theory 
Coupled to Gravity}\\
\vglue 0.5cm { Theodora Ioannidou$^\dagger${\footnote{{\it Email}:
ti3@auth.gr}} and
Burkhard Kleihaus $^\ddagger$
{\footnote{{\it Email}: kleihaus@theorie.physik.uni-oldenburg.de}}
} \\
\vglue 0.3cm
$^\dagger${\it Mathematics Division, School of Technology, 
Aristotle University of
Thessaloniki, Thessaloniki 54124,  Greece}\\
$^\ddagger${\it Institut f\"ur Physik, Universit\"at Oldenburg, Postfach 2503
D-26111 Oldenburg, Germany\\}
\end{center}

{\bf Abstract:} 
We construct non-BPS regular and black hole solutions of ${\cal N}=4$ $SU(N)$
supersymmetric Yang-Mills theory coupled to Einstein gravity. 
Our numerical studies reveal a number of interesting phenomena when
the gravitational constant $\al=M_{\rm YM}/gM_{\rm Planck}$ (where
$M_{\rm Planck}$ is the Planck mass and $M_{\rm YM}$ is the monopole mass) is
either weak (flat limit) or comparable to the Yang-Mills interaction.
In fact, black hole solutions exist  in a certain bounded domain in the 
$(\al, r_{\rm H})$ plane where $r_{\rm H}$ denotes the radius of the black
hole horizon.

\section{Introduction}

The $(3+1)$-dimensional ${\cal N}=4$ $SU(N)$ supersymmetric Yang-Mills theory
broken to $U(1)^{N-1}$ can be studied as an effective field theory on $N$ 
parallel D3-branes \cite{Wi1}.
The existence of three strings junction connecting 3 parallel D3-branes
 was first conjectured in \cite{Sc}. 
Subsequently, in  \cite{Hash} planar three strings junction have been found in 
explicit form. 
These states preserve $1/4$ of the supersymmetries
and correspond to static BPS spherically symmetric
solutions of  $SU(3)$ Yang-Mills-Higgs theory.
In \cite{IS1}, a class of static non-BPS dyon solutions of 
the aforementioned model has been derived which 
describes non-planar string junctions connecting $N$ D3-branes.
The solutions are spherical symmetric with electric charge determined by 
the Higgs vacuum expectations values (vevs).

In this paper, we extend this work by coupling the model with the 
Einstein gravity and construct the corresponding gravitating globally regular
and black hole solutions.
In the latter case, the strings meet at the horizon of the black hole
and therefore, the string spectrum does not consists of  junctions.
Recall that, black holes that are solutions of supersymmetric theories 
can be associated with solitons \cite{G}.

From a complementary point of view one expects to obtain these  BPS and non-BPS
solutions from the supergravity equations of motion for any BPS and non-BPS 
state in the spectrum.
The simplest BPS solutions of this kind are spherically symmetric black hole
solutions of ${\cal N}=2$  theories \cite{FKS} and there existence 
strongly depends on the  value of  the charges and vacuum moduli \cite{M}. 
Non-BPS composites could also exist \cite{D};
 however this is still an open question.

\section{The Model}
We consider a model in $(3+1)$ dimensions which consists of the  
Einstein-Hilbert action and the ${\cal N}=4$ $SU(N)$  supersymmetric 
Yang-Mills action \cite{eugen}, 
\bea
S=\int \left(\fr{1}{16\pi G}R+\mbox{tr}\left\{\kk_1F_{\mu\nu}F^{\mu\nu}+\kk_2
\sum_{I=1}^6 D_\mu\Phi^I D^\mu \Phi^I+\kk_3
\sum_{I,J=1}^6 [\Phi^I,\Phi^J]^2\right\}\right) \sqrt{-g}\, d^4x.
\eea
Here $\Phi^I$, $I=1,..,6$  denote the six Higgs scalars, 
with covariant derivatives defined by: 
$D_\mu\Phi^I=\partial_\mu\Phi^I-i[A_\mu,\Phi^I]$, 
$F_{\mu\nu}=\pr_\mu A_\nu-\pr_\nu A_\mu-i[A_\mu,A_\nu]$,
$R$ is the scalar curvature, and $g$ is the determinant of the metric.
$G$ denotes the gravitational coupling parameter and 
the constants $\kappa_i$ are fixed as: $\kappa_1=-\frac{1}{2}$,
$\kappa_2=-1$, $\kappa_3=\frac{1}{2}$.
In flat space this model can be considered as a dimensional reduction of the 
$(4+n)$-dimensional  Yang-Mills theory, where $n$ is the number
of the extra dimensions and is (also) equal to the number  of the Higgs fields
of the $(3+1)$-dimensional ${\cal N}=4$ $SU(N)$  supersymmetric 
Yang-Mills model.

For simplicity, we  use  the coordinates $r,z,\zb$ on $\R^3$   where in 
terms of the usual spherical coordinates $r,\theta,\phi$ the Riemann sphere 
variable $z$ is given by $z=e^{i\phi} \tan(\theta/2)$. 
Then the  Schwarzschild-like metric becomes
\begin{equation}
\label{metric}
ds^2=-A^2(r)B(r)\,dt^2+\fr{1}{B(r)}\,dr^2+\frac{4r^2}{(1+|z|^2)^2}
\,dz d\bar{z}, \hs B(r)=1-\frac{2m(r)}{r},
 \label{s}
\end{equation}
where  $A(r)$ and $B(r)$ are  real functions 
and depend only on the radial coordinate $r$, and
$m(r)$ is the mass function.
 The (dimensionfull) mass of the solution
is  $m_{\infty}\equiv m(\infty)$ and the square-root of the 
determinant is
\begin{equation}
\sqrt{-g}=iA(r)\,\frac{2r^2}{(1+|z|^2)^2}.
\end{equation}
Also, the Einstein equations  simplify to:
 \begin{equation}
 \frac{2}{r^2}\,m' = -8\pi
 G\, T^0_0,\hs  \hs
 \fr{2}{r} \fr{A'}{A}\,B =
-8\pi G \,\left( T^0_0-T^r_r \right)
\label{ff}
 \end{equation}
where  prime denotes the derivative with respect to $r$ and $T_{\mu\nu}=
g_{\mu\nu} {\cal L}_M-2(\pr {\cal L}_M/g^{\mu\nu})$ is
\bea 
T_{\mu\nu} & = & 
\tr\left(\kk_1(g_{\mu\nu} F_{\al\bt}F^{\al\bt}-4g^{\al\bt}F_{\mu\al}
 F_{\nu\bt})
\right) \nonumber \\ 
& &
+\!\kk_2 \sum_{I=1}^6\left(-2 D_\mu\Phi^I D_\nu\Phi^I+g_{\mu\nu}D_\al
\ \Phi^ID^\al \Phi^I
\right)  
+\kk_3g_{\mu\nu}\sum_{I,J=1}^6[\Phi^I,\Phi^J]^2 \ .
\eea

The harmonic map ansatz to obtain $SU(N)$ dyons \cite{IS1}
is of the form
\be
\Phi^I=\sum_{j=0}^{N-2} \beta^I_j(P_j-\frac{1}{N}), \ \ \ 
A_0=\sum_{j=0}^{N-2} \delta_j(P_j-\frac{1}{N}), \ \ \ 
A_z=i\sum_{j=0}^{N-2} \gamma_j[P_j,\partial_z P_j], \ \ \ 
A_r=0.
\label{ansatz}
\ee
Here $\beta^I_j(r),\gamma_j(r),\delta_j(r)$ are real functions depending only
 on the radial
coordinate $r$, and $P_j(z,\zb)$ are $N\times N$ hermitian projectors
 independent of $r$ and orthogonal i.e. $P_iP_j=0$
for $i\neq j.$
Note that we are working in a real gauge, so that $A_\zb=A_z^\dagger$.

The orthogonality of the projectors $P_j$ means that the Higgs fields
$\Phi^I$ are mutually commuting, i.e. $[\Phi^I,\Phi^J]=0,$ 
 so they
are simultaneously diagonalizable and this allows the eigenvalues to
 be interpreted as the positions of the strings in the transverse space.

The explicit form of the projectors is given as follows.
Let $f$ be the holomorphic vector
\be
f=(f_0,...,f_j,...,f_{N-1})^t, \ \ \mbox{where} \ \ f_j=z^j\sqrt{{N-1}\choose 
j}
\label{smap}
\ee
and ${N-1}\choose j$ denote the binomial coefficients.
Define the operator $\Delta$, acting on a vector 
$h\in \C^N$ as
\be
\pp h=\pr_z h- \fr{h \,(h^\dg \,\pr_z h)}{|h|^2}
\ee
then $P_j$ is defined as
\be
P_j=\frac{(\pp^j f)(\pp^j f)^\dagger}{\vert \pp^j f\vert^2}.
\ee
The particular form of these projectors corresponds to the requirement that
the associated dyons are spherically symmetric (see \cite{Za} for more
details). 

It is convenient to make a change of
variables to the following linear combinations
\be
\beta^I_j=\sum_{k=j}^{N-2}b^I_k, \hs c_j=1-\gamma_j-\gamma_{j+1},
 \hs
\delta_j=\sum_{k=j}^{N-2}d_k, \hs\mbox{for\ \ } j=0,\ldots,N-2
\label{change}
\ee
where we have defined $\gamma_{N-1}=0.$ 

The magnetic charges, $n_k$, for $k=1,..,N-1$, can be read off from the 
large $r$
behaviour of the magnetic field
\bea
B_i&=&\fr{1}{2}\,\varepsilon_{ijk}\, F_{ij}\nonumber\\
&\sim &\frac{\widehat x_i}{2 r^2}G
\eea
where $G$ is in the gauge orbit of 
\be G_0=\mbox{diag}(n_1,
\,n_2-n_1,\,\dots,\,n_{N-1}-n_{N-2},\,-n_{N-1}).
\ee
In the case of maximal symmetry breaking, which we shall consider here,
 they are given
by \cite{IS1}
\be
n_k=k(N-k), \ \ \ \ \ k=1,...,N-1.
\label{charges}
\ee
Similarly, the large $r$ asymptotic of the electric field
\be
E_i
\sim\frac{\widehat x_i}{2 r^2}A_0
\label{defg}
\ee
allows the electric charges (which classically are real-valued)
to be found from 
\be
A_0=\sum_{j=0}^{N-2} 2(P_j-\frac{1}{N}) (r^2{\delta_j}')\vert_{r=\infty}
\ee
ie. the electric charges are related to the $1/r$ coefficients of 
$\delta_j$ in a large $r$ expansion.

After some algebra, it can be shown  that
substitution of (\ref{ansatz}) for  $N=3$
 into  the energy-momentum tensor leads to
\bea
T^0_0-T^r_r&=&\fr{8\kk_1}{BA^2r^2}
\left(c_0^2d_0^2+c_1^2d_1^2\right)
+\fr{8\kk_1B}{r^2}\left(c_0'^2+c_1'^2\right)+\fr{4\kk_2B}{3}
\sum_I\left((b^I_0)'^2+(b^I_1)'^2+(b^I_0)'(b^I_1)'\right)\nonumber\\
\label{t00-rr}\\
T^0_0&=&\fr{4\kk_1B}{r^2}\left(c_0'^2+c_1'^2\right)
+\fr{4\kk_1}{BA^2r^2}\left(c_0^2d_0^2+c_1^2d_1^2\right)
+\fr{2\kk_2}{r^2}\sum_I\left(c_0^2(b^I_0)^2+c_1^2(b^I_1)^2\right)
\nonumber\\
&&+\fr{4\kk_1}{3A^2}\left(d_0'^2+d_1'^2+d_0'd_1'\right)
+\fr{2\kk_2B}{3}\sum_I\left((b^I_0)'^2+(b^I_1)'^2+(b^I_0)'(b^I_1)'\right)
\nonumber\\
&&+\fr{4\kk_1}{r^4}
\left[(1-c_0^2)^2+(1-c_1^2)^2-(1-c_0^2)(1-c_1^2)\right], \label{trr}
\eea
while variation of the energy density  with respect to  matter fields
leads to the following system of ODE's
\bea
\fr{1}{A}\left(ABr^2 \,{b^I_0}'\right)'&=&
2\left(2c_0^2 b^I_0-c_1^2 b^I_1\right)
\nonumber\\
\fr{1}{A}\left(ABr^2\,{b^I_1}'\right)'&=&
2\left(2c_1^2 b^I_1-c_0^2 b^I_0\right)
\label{b1}\\
BA\left(\fr{r^2}{A} \, {d_0}'\right)'&=& 
2\left(2c_0^2 d_0-c_1^2 d_1\right)
\nonumber\\
BA\left(\fr{r^2}{A} \, {d_1}'\right)'&=& 
2\left(2c_1^2 d_1-c_0^2 d_0\right)\
\label{d1}\\
\fr{1}{A}\left(AB\,c_0'\right)'&=&c_0
\left[\fr{1}{r^2}(2c_0^2-c_1^2-1)+\fr{1}{BA^2}\,d_0^2
+\fr{\kk_2}{2\kk_1}\sum_I(b^I_0)^2\right]
\nonumber\\
\fr{1}{A}\left(AB\,c_1'\right)'&=&
c_1\left[\fr{1}{r^2}(2c_1^2-c_0^2-1)+
\fr{1}{BA^2}\,d_1^2
+\fr{\kk_2}{2\kk_1}\sum_I(b^I_0)^2\right],
\label{c1}
\eea
with the appropriate boundary conditions due to the finiteness of the
energy density.

Since we restrict to  maximal symmetry breaking 
$SU(3)\rightarrow U(1)^2$,
(ie. to three D3-branes being distinct in transverse space),
the boundary conditions on $c_j(r)$ are: 
$c_j(\infty)=0$ where $j=0,1$.
The remaining free parameters: $b^I_j(\infty)$,
 determine the vevs of the Higgs scalars since
the ansatz (\ref{ansatz}) along the positive $x_3$-axis (that is, 
by setting $z=0$) under the change of variables 
(\ref{change}), results in
\bea
\Phi^I(r) & = & \frac{1}{3}\mbox{diag}(2b^I_0+b^I_1,-b^I_0
+b^I_1,-b^I_0-2b^I_1)
\label{vevs}
\eea
from which the Higgs vevs can be read off in terms of $b^I_j(\infty).$ 

By writing the components of (\ref{vevs}) as
\be
\Phi^I(r)=\mbox{diag}(\Phi^I_1(r),\Phi^I_2(r),\Phi^I_3(r))
\ee
the positions of the three D3-branes in the two-dimensional 
transverse space are given by
\be
(x^4_\alpha,x^5_\alpha)=(\Phi^1_\alpha(\infty),\Phi^2_\alpha(\infty)) 
\hs \mbox{for} \hs \alpha=1,2,3,
\label{pos}
\ee
while their values  for different $r$ correspond to
 the positions of the strings
which form the string junction and end on the D3-branes.
 (More details  in section 3).

For globally regular solutions the boundary conditions of the matter profile functions are:
\bea
d_i(r=0)=0,\hs  c_i(r=0)=1, \hs b^I_i(r=0)=0, \hs m(r=0)=0, \hs i=0,1,
\label{bc_org}
\eea
and describe a string junction  formed at the origin.
However, black holes  possess an event horizon 
at $r = r_{\rm H}$  determined by: $B(r_{\rm H})=0$, 
or (equivalently) by:  $m(r_{\rm H})=r_{\rm H}/2$.
The horizon radius is  a singular point of the differential equations and so,
regularity of the solutions (due to (\ref{t00-rr})), impose 
the following boundary conditions:
\be
d_i(r_{\rm H})=0, \hs i=0,1.
\ee
In what follows  we  consider only purely magnetic solutions ie. $A_0=0$
which implies that $d_i(r)=0$, for $i=0,1$.
Then eqs.~(\ref{b1}) and (\ref{c1}) evaluated at $r=r_{\rm H}$  yield
\bea
\left[2\left(2c_0^2 b^I_0-c_1^2 b^I_1\right)-r^2 B'{b^I_0}'\right]_{\rm H}
 & = & 0 
\nonumber\\
\left[2\left(2c_1^2 b^I_1-c_0^2 b^I_0\right)-r^2 B'{b^I_1}'\right]_{\rm H} 
& = & 0
\label{bc_b}\\
\left[
c_0\left\{\fr{1}{r^2}(2c_0^2-c_1^2-1)
+\fr{\kk_2}{2\kk_1}\sum_I(b_0^I)^2\right\}- B' c_0'
\right]_{\rm H} & = & 0
\nonumber\\
\left[
c_1\left\{\fr{1}{r^2}(2c_1^2-c_0^2-1)
+\fr{\kk_2}{2\kk_1}\sum_I(b_1^I)^2\right\}- B' c_1'
\right]_{\rm H} & = & 0,
\label{bc_c}
\eea
respectively.

\section{Numerical Simulations}

Next we study the deformation of  the classical soliton
solutions of ${\cal N}=4$ supersymmetric Yang-Mills equations 
to gravitating ones and black holes.
In particular, we investigate the deformations as the gravitational
parameter $\alpha=\sqrt{4\pi G}$ goes from zero to its maximum 
value. In this paper the asymptotic values of the Higgs field 
are fixed, ie.
\be
b_0^1=-3/4,\hs b_1^1=1/4, \hs b_0^2=b_1^2=-1/3.
\label{av}
\ee

\subsection{Globally Regular Solutions}

Globally regular solutions describe self-gravitating monopoles.
A heuristic argument \cite{gmono,BFM} suggests that they 
cannot persist if the coupling to gravity becomes too strong; ie.
 for  $\alpha$ of order one. 

In particular, as   $\alpha$ departs from zero
a first branch of solutions emerges from the flat space monopole.
When 
$\al$ reaches its maximal value: $\alpha_{\rm max}\approx 1.349$
the first branch merges with a second one which bends back  
to a critical value $\alpha_{\rm cr}\approx 1.345$.
The mass of the second branch is slightly larger than of
 the first one, indicating an instability.
On the second branch, when  $\alpha \ra \alpha_{\rm cr}$ 
   the minimum  of $B(r)$ tends to zero and so a
degenerate horizon is formed at: $r_{\rm deg}=2 \alpha_{\rm cr}$.
In this limit the  solution consists of an abelian part in 
the outside region $r_{\rm deg}<r<\infty$,   
where the metric  
coincides with that of the extreme Reissner-Nordstr\"om
black hole of charge two, the gauge functions $c_0(r)$ and $c_1(r)$
vanish identically and the Higgs functions $b^I_0(r)$ and  $b^I_1(r)$ 
are constant; 
and of a non-abelian part
in the inside region $0\leq r<r_{\rm deg}$,
where the functions interpolate
continuously between their values at the origin and at $r_{\rm deg}$.
As in \cite{gmono,BFM}, the appearance of a double zero of 
$B$ implies that the degenerate horizon is at
an infinite physical distance from the origin, which means that the  monopole
is located inside the degenerate horizon.

These results are similar to the gravitating monopoles of 
$SU(2)$ Einstein-Yang-Mills-Higgs theory which have been discussed extensively
in \cite{gmono,BFM}.
In Fig.~1  we plot  the strings in the $x^4-x^5$ plane
for several values of  $\alpha$ and observe
that they do not change considerably as $\alpha$ increases from 
zero to one. Even for greater values of $\alpha$ the strings are deformed
only slightly.

\begin{figure}[h!]
\parbox{\textwidth}{
\centerline{
{\epsfysize=12.0cm \epsffile{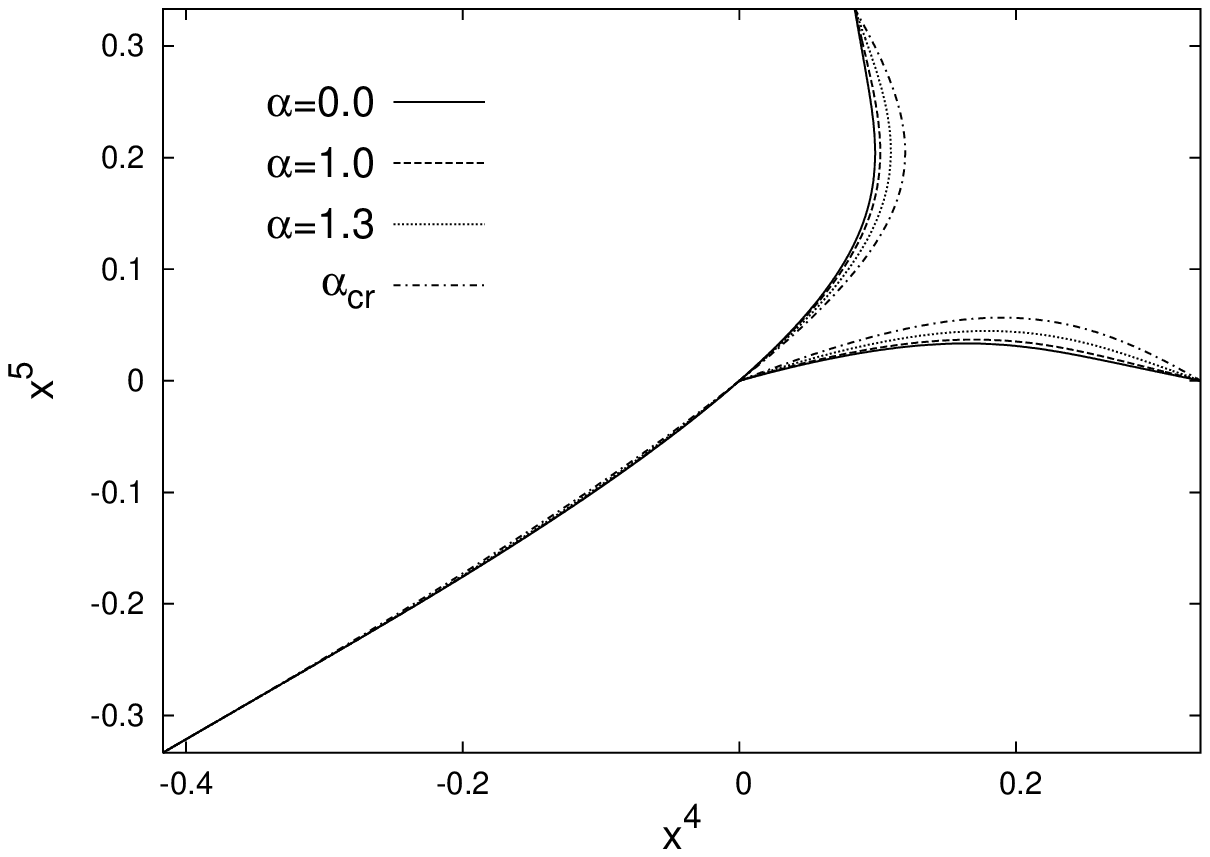} }
}\vspace{0.5cm}
{\bf Fig.~1} \small
Three strings junction with boundary conditions given by (\ref{av})
 for several  values of the gravitational coupling parameter.
\vspace{0.5cm}
}
\end{figure}

\subsection{Black Hole Solutions}

Next we study the construction of black hole solutions where 
three special cases arise. Case I describes
embedded Abelian solutions and  occurs when 
$c_0(r)= c_1(r) = 0$; Case II describes 
embedded non-Abelian solutions with gauge group $SU(2)\times U(1)$
and occurs when either 
$c_0(r) = 0$, $c_1(r) \neq 0$ (IIa) or
$c_0(r) \neq 0$, $c_1(r) = 0$ (IIb); and Case III
describes genuine non-Abelian $SU(3)$ solutions and occurs when 
both $c_0(r)$ and  $c_1(r)$ are different from zero.

The numerical simulations show that:
  in case I and II the presence of Abelian gauge fields 
impose a lower bound 
in the size of the black holes which  can not be arbitrarily small;
while in case II and III black holes cannot become arbitrarily large
since (heuristically) their radius 
cannot exceed the radius of the non-Abelian core of the monopole.

\subsubsection{Case I}
When  $c_0(r)=c_1(r) = 0$ the Higgs profile functions are constant  
$b_0^I(r) = b_0^I(\infty)$, $b_1^I(r) = b_1^I(\infty)$ while
 the metric ones become
\bea
m(r) & = & m_\infty - \fr{\alpha^2}{2} \fr{4}{r},
\nonumber \\
A(r) & = & 1.
\label{RN}
\eea
where $m_\infty$ is an integration constant.
The solution corresponds to the Reissner-Nordstr\"om one 
with mass $m_\infty$, charge 
two and  event horizon given by:
$r_{\rm H} =m_\infty + \sqrt{ m_\infty^2 - 4 \alpha^2}$.
Note that, the black holes exist for  
$r_{\rm H} \geq 2 \alpha$ and the equality holds for the extremal case.
Fig.~2 presents the domain of existence of the black holes
 in the $\alpha-r_{\rm H}$ plane and
indicates that  charge two Reissner-Nordstr\"om black holes exist for 
any   $\alpha$ and $r_{\rm H}$ above the 
thick dashed line $r_{\rm H} = 2\alpha$.

\begin{figure}[h!]
\parbox{\textwidth}{
\centerline{
{\epsfysize=12.0cm \epsffile{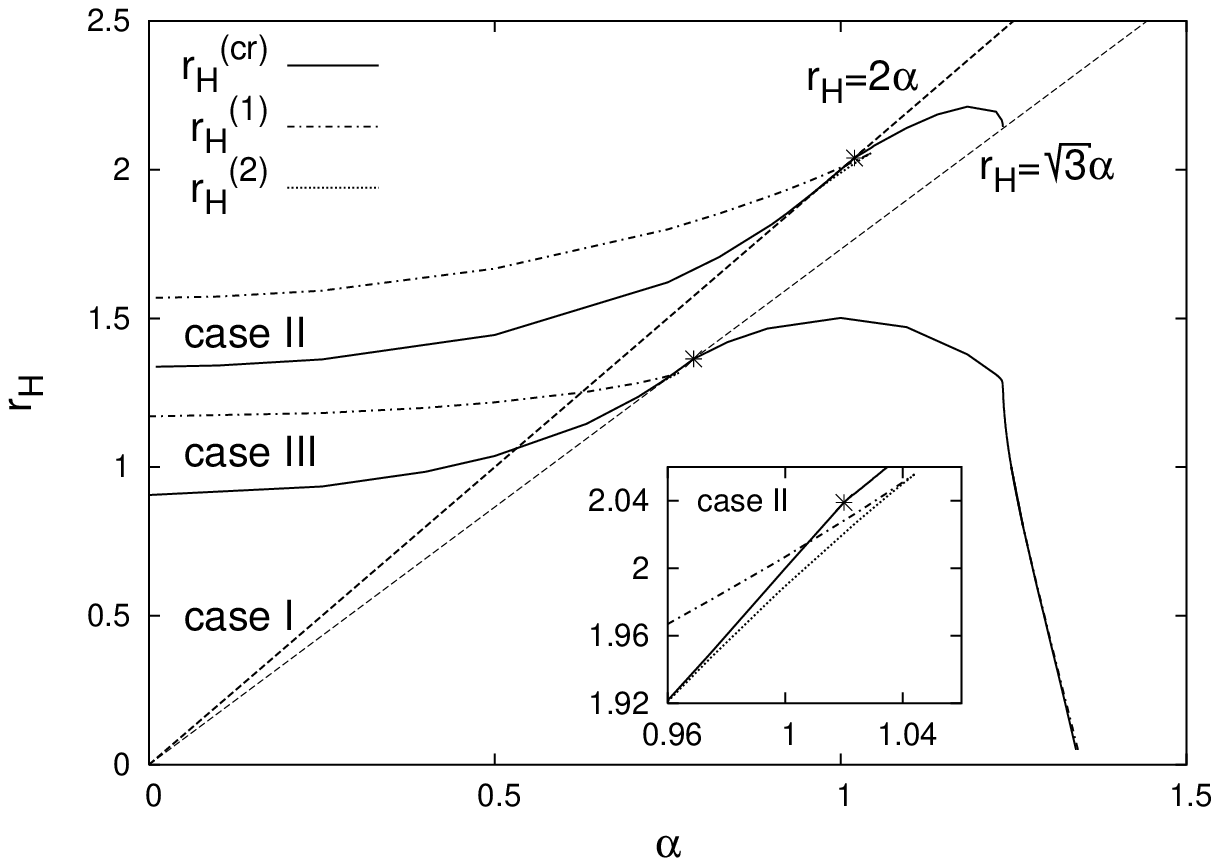} }}\vspace{0.5cm}
{\bf Fig.~2} \small
The domain of existence of  black holes
with boundary conditions given by (\ref{av}).
\vspace{0.5cm}}
\end{figure}

\subsubsection{Case II}

Cases IIa and IIb are equivalent due to the symmetry of
 field equations under
the interchange $(c_0, b_0^I) \leftrightarrow (c_1, b_1^I)$,
(therefore, we concentrate on case IIa). 
When $c_0=0$ eqs.~(\ref{b1}) are combined to a single one
which after integration simplify to
\be
2{b^I_0}'+{b^I_1}' = \fr{\mbox{const}}{AB r^2}.
\ee
However, at the horizon $B(r_{\rm H})=0$ and thus
for regular solutions, the constant has to vanish ie.
\be
b^I_0(r) = -\fr{1}{2} b^I_1(r) + \fr{1}{2}\left[ 2 b^I_0(\infty) 
+ b^I_1(\infty)\right].
\label{II}
\ee
By setting $c_1(r) = w(r)/\sqrt{2}$ and $b^I_1(r)= h^I(r)$ 
the remaining differential  equations reduce to
\bea  
\left(ABr^2\,{h^I}'\right)' & = & 2 w^2 h^I A
\label{b1_q0_su21}\\
\left(AB\,w'\right)'
&=& 
w\left[\fr{1}{r^2}(w^2-1)+\sum_I(h^I)^2\right] A
\label{c1_q0_su21} \\
m' &=& \alpha^2 \left\{ B {w'}^2
+\sum_I w^2(h^I)^2 +\fr{B}{2}\sum_I r^2{(h^I)'}^2+\fr{1}{2r^2}
\left[(1-w^2)^2+3\right]\right\} \label{m_su21}\\
A'&=&\fr{2 \alpha^2}{r} A \left\{w'^2+\fr{1}{2}\sum_I r^2 {(h^I)'}^2\right\}
\label{A_su21} \ ,
\eea
which is the
Einstein-Yang-Mills-Higgs system with gauge group
$SU(2)\times U(1)$; multiple Higgs fields; and 
magnetic charge   $Q=\sqrt{3}$ (due to  the $U(1)$ field).
  
Next, we argue why   the black hole solutions of 
eqns.~(\ref{b1_q0_su21}-\ref{A_su21})
exist only if $r_{\rm H} \geq \sqrt{3} \alpha$,
as shown in  Fig.~2:
The constraint $B'(r_{\rm H}) \geq 0$ is true since otherwise 
$B(r)$ would be negative  near the horizon for $r>r_{\rm H}$
and  equal to  zero  at some point $r_0$
(due to the asymptotic value $B(\infty)\ra 1$).
But such a  point is a singular point of the 
equations of motion  and therefore, a smooth solution is  not guaranteed.
Applying  the above constraint  in  (\ref{m_su21})
gives the inequality
\be
1- \fr{3\alpha^2}{r_{\rm H}^2} \geq 
2 \alpha^2\left\{ \sum_I w_{\rm H}^2(h_{\rm H}^I)^2 
+\fr{1}{2r_{\rm H}^2}\left[(1-w_{\rm H}^2)^2\right]
\right\}.
\label{su2_ex_cond}
\ee
However, the right-hand side is non-negative 
implying that  $r_{\rm H}\geq \sqrt{3}\alpha$ and therefore,
no globally regular solutions exist in case II.
{\it Remark:} Note that in both cases I and II 
the lower bound of the horizon radius: $r_{\rm H}\geq Q\alpha$, is 
a consequence of the presence of an Abelian field.

In addition, in case II  an upper bound on black holes also exists.
Fig.~2 indicates that for   fixed $\alpha \leq 0.959$ and 
varying $r_{\rm H}$ a first  branch of solutions 
extends up to the maximal value  $r_{\rm H}^{(1)}(\alpha)$,
 where it merges with  a second branch which 
bends back to  the critical value 
$r_{\rm H}^{(cr)}(\alpha)$ and finally  bifurcates with  the abelian 
Reissner-Nordstr\"om solution of case I. 
The curves  $r_{\rm H}^{(1)}(\alpha)$ and 
$r_{\rm H}^{(cr)}(\alpha)$ are presented by the dash-dotted and solid lines, 
respectively, in Fig.~2.

Moreover, as shown in the inlet of Fig.~2, for $0.959<\al< 1.044$  three 
branches of solutions exist.
The first and second branch merge at 
$r_{\rm H}^{(1)}(\alpha)$, while the second and third branch 
merge at $r_{\rm H}^{(2)}(\alpha)<r_{\rm H}^{(1)}(\alpha)$ and
the third one finally  terminates at
$r_{\rm H}^{(cr)}(\alpha)$. 
The curves $r_{\rm H}^{(1)}(\alpha)$ and  $r_{\rm H}^{(2)}(\alpha)$ merge 
at $\alpha=1.044$ while  for larger values of $\alpha$ 
only one branch of solutions exists up to 
$r_{\rm H}^{(cr)}(\alpha)$.  
Note that, the upper  $r_{\rm H}^{(cr)}(\alpha)$ 
and  lower bound $r_{\rm H} = \sqrt{3} \alpha$ of the horizon radius coincide
 at  some value of the gravitational parameter 
$\alpha= \alpha^{\rm max}_{II} \approx 1.235$ above which 
no case II black hole solutions exist. 

At the critical  radius $r_{\rm H}^{(cr)}(\alpha)$ 
a bifurcation with the Reissner-Nordstr\"om solution of case I 
occurs only when  $\alpha \leq \alpha^*_{II}=1.02$.
For $\alpha > \alpha^*_{II}$, 
 the branches terminate since a degenerate horizon is formed.
In the limit $r_{\rm H}\ra r_{\rm H}^{(cr)}(\alpha)$  
the local minimum of the metric function $B$  
becomes equal to zero at $r_{\rm deg}=2\alpha$.
The limiting solution can be described as follows:
for $r>r_{\rm deg}$ the metric and gauge potential correspond to an
 extremal
abelian black hole with magnetic charge two and constant  Higgs field.
However, inside the degenerate horizon $r_{\rm H} \leq r<r_{\rm deg}$ 
a non-Abelian core persists.
The formation of the degenerate horizon is similar to 
the one of the  globally regular solutions, discussed in section {\bf 3.1}.

In order to understand the transition to Abelian black holes on a 
qualitative level,
one can consider the monopole as an extended object with a black hole
inside the non-Abelian core. 
When the   black hole becomes  larger than
the monopole core, the non-Abelian fields outside the horizon can not persist any
more and transform to Abelian ones.
The presence  of different
branches of solutions indicates the existence of instability, ie.
solutions with the largest masses are less stable.
Fig.~3 demonstrates the bifurcation with the abelian solution of case I 
and presents the mass of the solutions in all  cases
for $\alpha=0.5$.

\begin{figure}[h!]
\parbox{\textwidth}{
\centerline{
{\epsfysize=12.0cm \epsffile{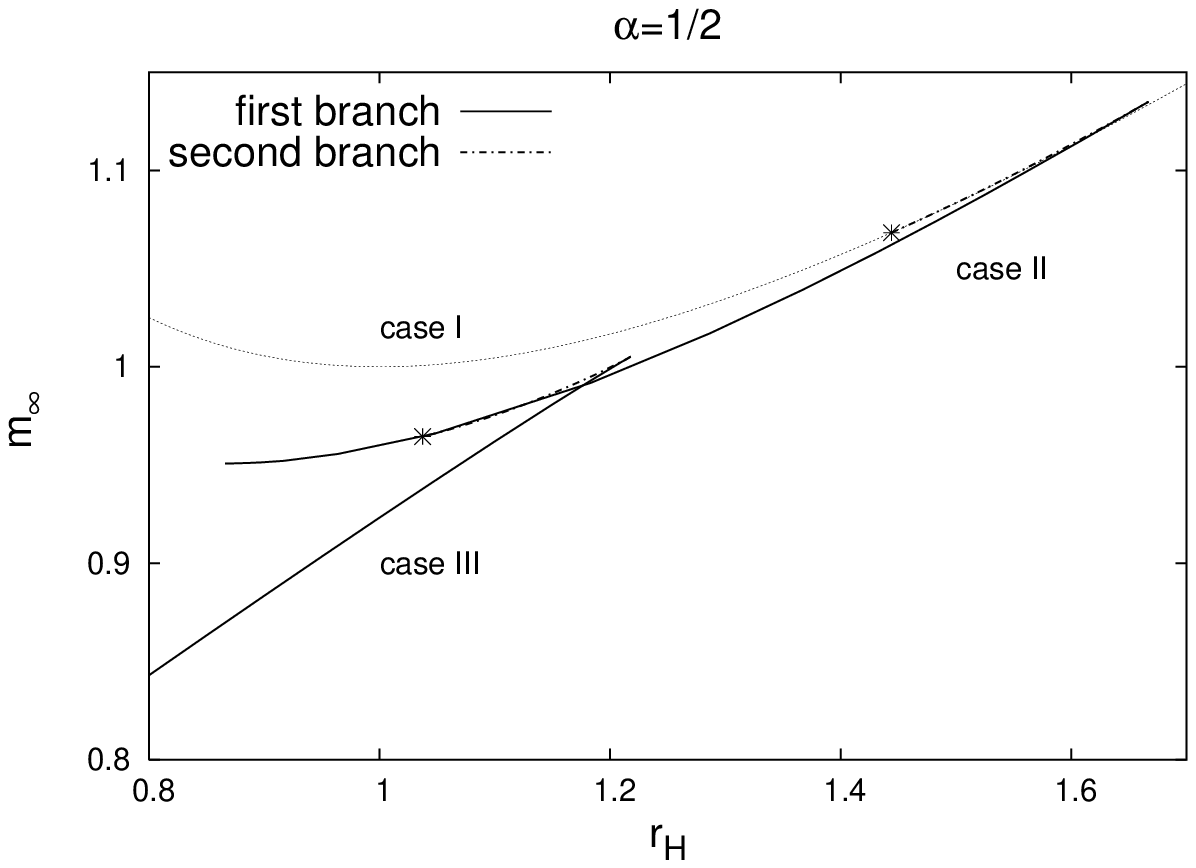} }
}\vspace{0.5cm}
{\bf Fig.~3} \small
The mass of the black hole solutions of case I,  II and  III as function of 
$r_{\rm H}$ for $\alpha=0.5$.
(Asterisk indicates the bifurcation point.)
\vspace{0.5cm}
}
\end{figure}

The functions $B(r)$, $A(r)$, $f(r)$, $h^1(r)$ and $h^2(r)$ are plotted 
 in Figs.~4a--4d for different values of the 
horizon radius  when $\alpha=\sqrt{1.3}$.
These functions are plotted  in the region $r \in [r_{\rm H},3]$ so that
the formation of the horizon is demonstrated.

\begin{figure}[h!]
\parbox{\textwidth}{
\centerline{
{\epsfysize=5.0cm \epsffile{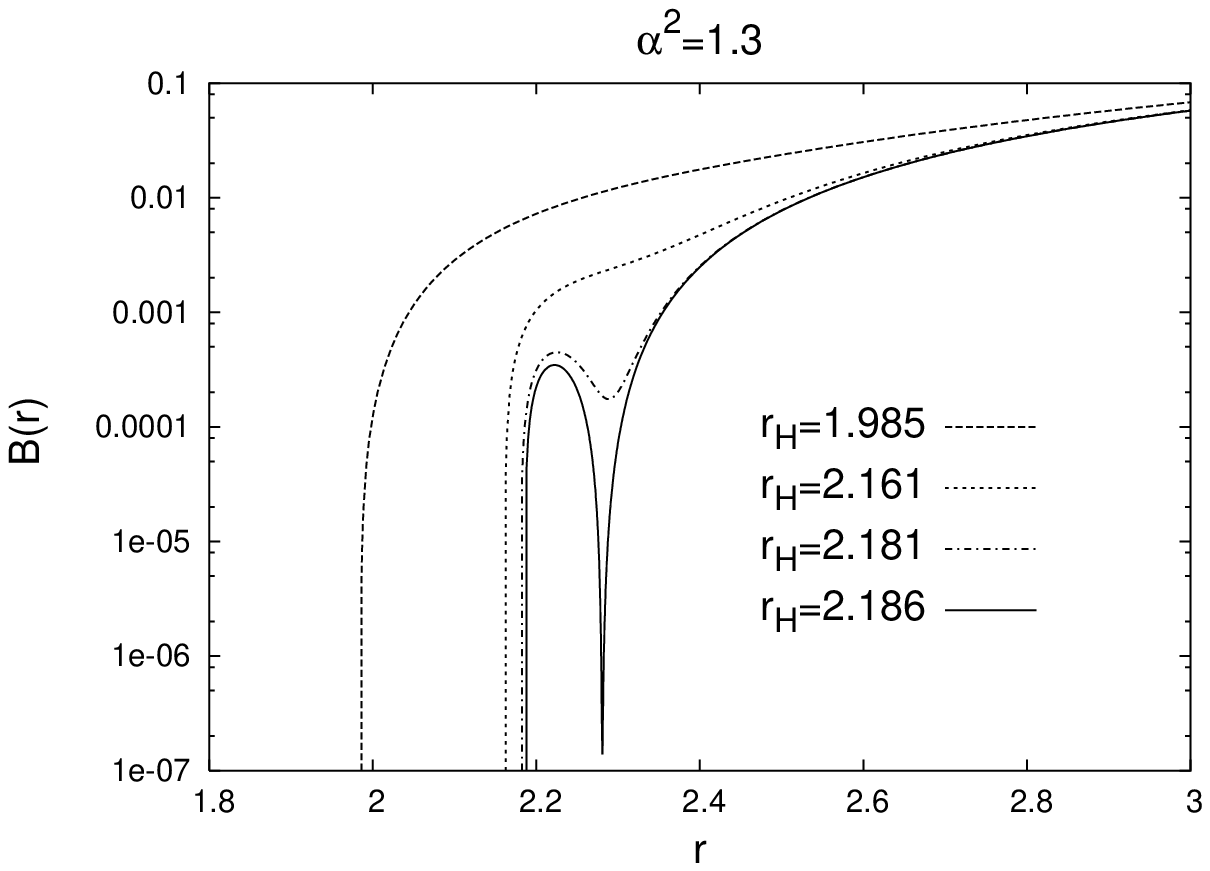} \hspace{0.5cm}\epsfysize=5.0cm 
\epsffile{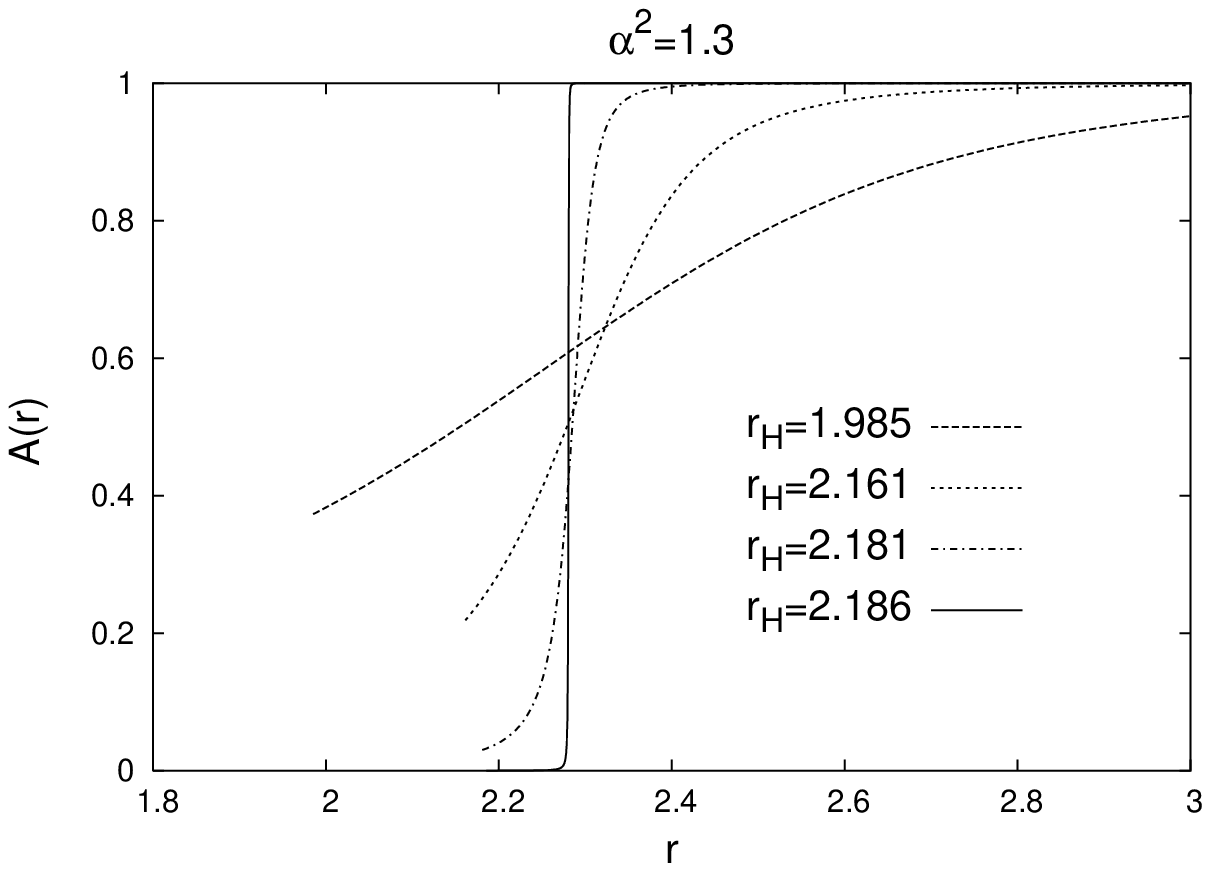}}
}\vspace{0.5cm}
\centerline{
{\epsfysize=5.0cm \epsffile{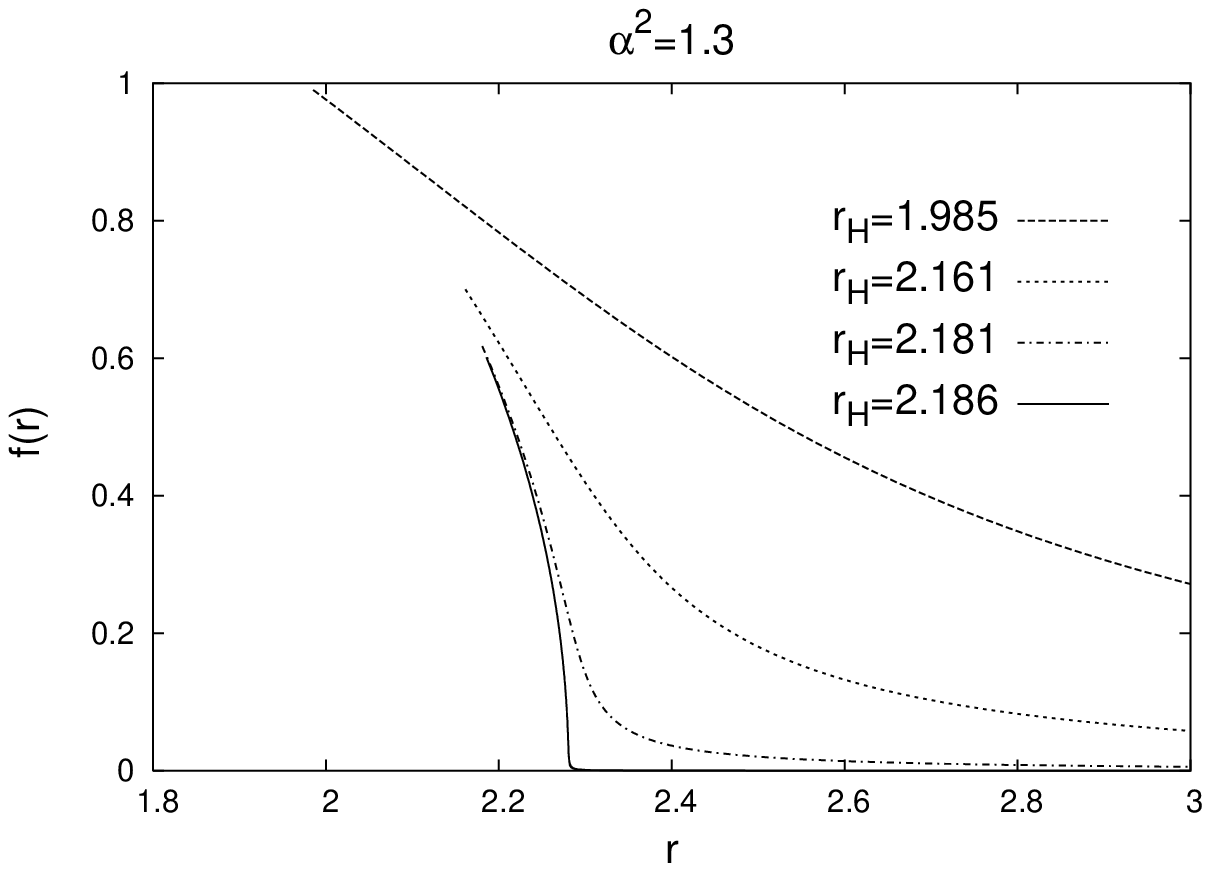} \hspace{0.5cm}\epsfysize=5.0cm 
\epsffile{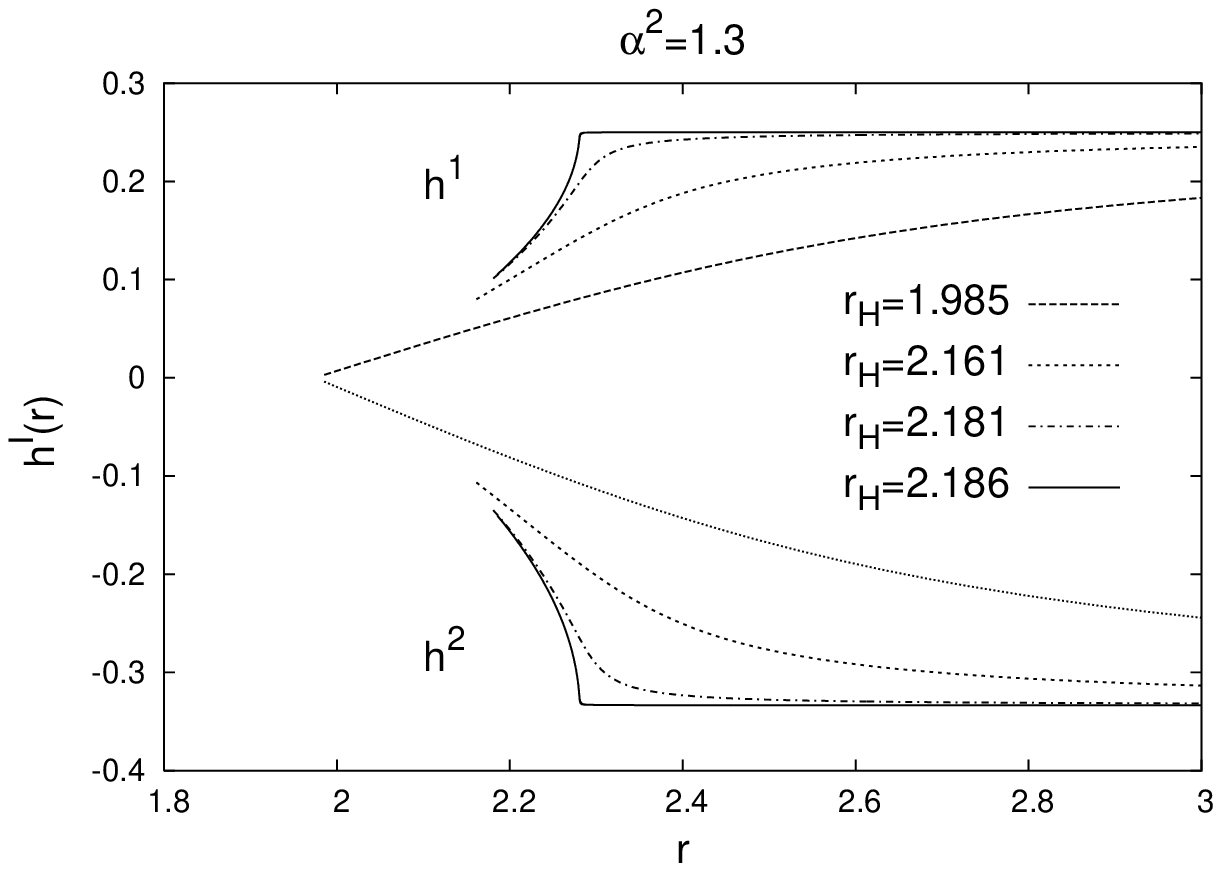}}
}\vspace{0.5cm}
{\bf Fig.~4} \small
The metric functions $B(r)$ (upper left), $A(r)$ (upper right), 
the gauge field function $f(r)$ (lower left) and the Higgs field 
functions $h^1(r)$, $h^2(r)$ (lower right)  in terms of $r_{\rm H}$ 
and $\alpha=\sqrt{1.3}$.}
\end{figure}

\subsubsection{Case III}

For the genuine $SU(3)$ black holes no Abelian gauge field is present and
therefore, no lower bound on the horizon radius exists. Indeed,
for vanishing horizon radius the black hole solutions tend
pointwise to the globally regular solutions.
However, as in case II, the solutions  bifurcate with  non-extremal 
black holes  when $\alpha$ is small; while  a degenerate 
horizon is formed for large values of $\alpha$ with transition point at:
$\alpha^*_{III} \approx 0.787$.

Let us consider  first the case $\alpha < \alpha^*_{III}$.
For $\alpha$ small, a first branch of black hole solutions emerges
from the  globally regular solutions with increasing horizon radius which
merges with a second branch  at the  maximal value $r_{\rm H}^{(1)}(\alpha)$, 
and then bends back to a critical value 
$r_{\rm H}^{(cr)}(\alpha)$. 
At this critical value the second branch bifurcates
with the first branch of case II  (in contrast with the
abelian solutions of case II),  as demonstrated 
in Fig.~3. 
Similarly with case II, three branches of solutions exist
 for $0.74 \leq \alpha \leq 0.766$ and  (only) one 
 for $0.766 \leq \alpha \leq \alpha^*_{III}$;
which bifurcate with the case II ones at the limit:
$r_{\rm H}\ra r_{\rm H}^{(cr)}(\alpha)$.

The transition to the $SU(2) \times U(1)$ black hole solutions
of case II  can be explained by the following argument:
the Higgs fields at infinity define two vector boson masses and core radii 
$m_0^2 = \left[ (b_0^1)^2+(b_0^2)^2 \right]_\infty$, $R_0=1/m_0$ and 
$m_1^2 = \left[ (b_1^1)^2+(b_1^2)^2 \right]_\infty$, $R_1=1/m_1$,
leading to the exponential decay of  $c_0$ and $c_1$, respectively.
Since $m_0 > m_1$ (due to (\ref{av})), $c_0$ is essentially zero outside 
$R_0$   where $SU(3)$ breaks to $SU(2) \times U(1)$.
Consequently, if the horizon radius is of  the order of $R_0$, 
an $SU(2)$ gauge field and  a $U(1)$ field exists outside the horizon.

For  $\alpha > \alpha^*_{III}$  two distinguished 
cases exist when either   $\alpha \leq \alpha^{\rm max}_{II}$ or 
$\alpha \geq \alpha^{\rm max}_{II}$. 
In the first case,  only one branch of solutions exists which terminates at 
$r_{\rm H}=r_{\rm H}^{(cr)}(\alpha)$ and  a degenerate horizon forms 
at $r_{\rm deg}= \sqrt{3}\alpha$.
The limiting solution coincides with 
the extremal case II solution in the outside region  $r>\sqrt{3}\alpha$
when:  $c_0(r)=0$, $c_1(r)=w(r)$, $2 b^I_0(r) +b^I_1(r)=\mbox{const}$, 
$b^I_1(r)=h^I(r)$ and  $B(r)=A(r)$.
In the inside region $r<\sqrt{3}\alpha$, the functions $c_0(r)$ and 
$2 b^I_0(r) +b^I_1(r)$ are non-trivial; and  the situation is 
 similar to the solutions of case II when
 the extremal abelian solution is replaced by the extremal case II solution.
Thus, we see again that a horizon is  formed at $r_{\rm deg}$
 when  the monopole core becomes too massive.
 However, for the genuine SU(3) solutions
there is (still) a non-Abelian gauge field outside $r_{\rm deg}$ due
 to the lighter mass $m_1$.

In the second case the extremal case II solutions do not exist since 
$\alpha >  \alpha^{\rm max}_{II}$.
In fact as the horizon radius tends to $r_{\rm H}^{(cr)}(\alpha)$,
a degenerate horizon forms at $r_{\rm deg}=2\alpha$.
In this limit the solution in the outside region is 
Abelian with magnetic charge two; while
 in the inside region  is non-Abelian (as case II).
For larger values of $\alpha$, two branches of solutions exist which m
erge at some $r_{\rm H}^{(1)}(\alpha)$; while a degenerate horizon forms
on the second branch as $r_{\rm H}\ra r_{\rm H}^{(cr)}(\alpha)$.
There are no genuine non-Abelian black hole solutions for
 $\alpha>\alpha_{\rm max}$.

\section{String Junction}

Recall that the vevs of the Higgs fields determine the position
 of the strings in transverse space due to (\ref{pos}).
For globally regular solutions the Higgs functions vanish at the origin
and the three strings are joint  in transverse space.
However, in black holes  the origin is replaced by their
horizon and thus the strings are not connected in general.

Next we discuss the string interpretation of our  solutions in all
three cases:

Case I is the simplest one since the  symmetry breaking is $U(1)^2$.
The Higgs field equals its vacuum value everywhere in space
and so the strings  degenerate to a point on  the brane.

In case II, the symmetry breaking is $SU(2)\times U(1)$ and thus
 the Higgs functions are linearly dependent as shown in 
(\ref{II}). So the first string degenerates to the  point
\be
(x^4_1, x^5_1)=\left(\fr{1}{3}\left( 2 b^1_0(\infty) + b^1_1(\infty)\right),
\fr{1}{3}\left( 2 b^2_0(\infty) + b^2_1(\infty)\right) \right) 
\ee
since one of the branes decouples
when the non-Abelian symmetry group is reduced from $SU(3)$ to $SU(2)$.
Recall that  in case IIa, the Higgs profile functions are related since
\be
 h^1(r)/h^1(\infty) = h^2(r)/h^2(\infty)
\ee
which implies that $x^5_a$ is a linear function of $x^4_a$, for $a=2,3$. 
Therefore, the strings 2 and 3 form straight lines in transverse space 
connecting the branes at 
$\left(x^4_a(\infty),x^5_a(\infty)\right)$ to the point 
$\left(x^4_a(r_{\rm H}),x^5_a(r_{\rm H})\right)$.

Fig.~5 presents the string coordinates at the horizon 
$(x^4_i,x^5_i)|_{r_{\rm H}}$ 
(for $i=1,2$) as functions of  $r_{\rm H}$ 
when  $\alpha= 1.225, 1, 0.75$
and shows that the strings are not joint in the non-extremal case since
 $\left(x^4_2,x^5_2\right)|_{r_{\rm H}} \neq 
\left(x^4_3,x^5_3\right)|_{r_{\rm H}}$ for $r_{\rm H} > \sqrt{3} \alpha$.
However, in the extremal case $r_{\rm H} = \sqrt{3} \alpha$ 
the  strings 2 and 3 are jointed at the horizon since 
$b_1^I(r_{\rm H})=0$.

Next we investigate the consequences of the transition to Abelian black 
holes when $\alpha \leq \alpha^*_{II}$ and the formation of a
 degenerate horizon when $\alpha > \alpha^*_{II}$.
 Fig.~5 shows  that  for $\alpha \leq \alpha^*_{II}$
the coordinates $x^4_{2,{\rm H}}$, $x^5_{3,{\rm H}}$ increase 
monotonously with $r_{\rm H}$ up to a maximal value
where the transition to Abelian black holes occurs.
 [Similarly, $x^5_{2,{\rm H}}$, $x^4_{3,{\rm H}}$
decrease monotonously]. 
At the transition point the coordinates 
$(x^4_{a,{\rm H}},x^5_{a,{\rm H}})$ are equal to 
$(x^4_a(\infty),x^5_a(\infty))$ --- hence the strings 
degenerate to points on the branes.
Non-degenerate strings cannot exist for black holes larger than 
the non-Abelian core.
For $\alpha > \alpha^*_{II}$ a degenerate horizon 
forms outside the non-Abelian core when $r_{\rm H}\ra r^{(cr)}_{\rm H}$
 and  the strings stretch between the (event) horizon
and the degenerate one.

\begin{figure}[h!]
\parbox{\textwidth}{
\centerline{
{\epsfysize=12.0cm \epsffile{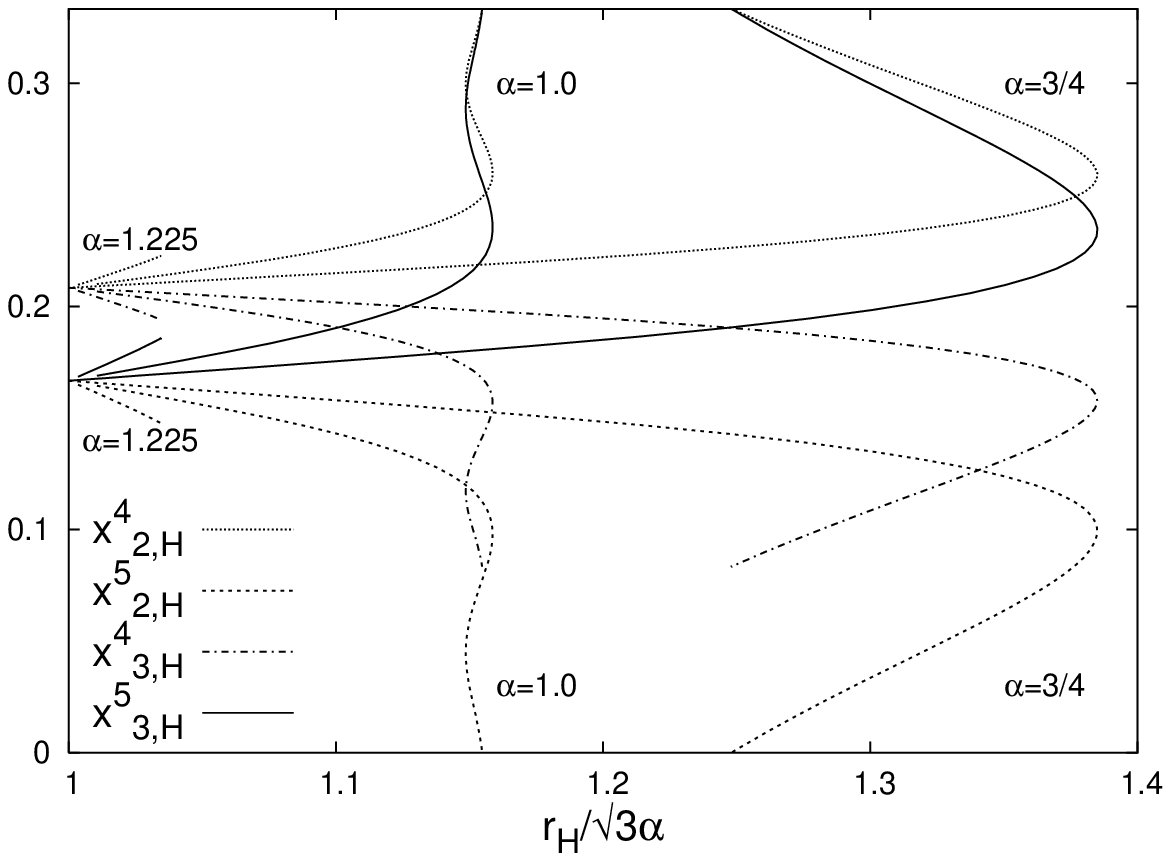} }
}\vspace{0.5cm}
{\bf Fig.~5} \small
The coordinates of the  strings 2 and 3 at the horizon as function of the
scaled horizon radius $r_{\rm H}/\sqrt{3}\alpha$  for $\alpha=1.225,1, 3/4$.
\vspace{0.5cm}}
\end{figure}

In case III, the strings do not degenerated at a point as shown in 
Fig.~6 where the  $x^4-x^5$ plane for $r_H=0.1, 1$ 
and $\alpha=1.0$ is plotted.
 However,  the separation  of the end points at the horizon 
gets smaller for  small horizon radius,
and the strings  joint as the black hole solutions approach
the globally regular solutions (in the limit of a vanishing horizon radius).
Similarly to case II, for $\alpha \leq \alpha^*_{III}$ the 
$SU(3)$ black holes bifurcate with
non-extremal $SU(2)\times U(1)$ black holes which implies that 
one string degenerates
to a point, while the others are associated with the gauge group $SU(2)$.
On the other hand, for $\alpha^*_{III}< \alpha \leq \alpha^{\rm max}_{II}$, 
a degenerate horizon forms at $r_{\rm deg}$ leaving three non-degenerate 
strings in the inside and two strings in the outside region as
demonstrated in Fig.~6.
[Note that, the existence of the string in the outside region
is due to the presence of the non-Abelian $SU(2)$ gauge fields].
In fact, since (in the outside region) the solutions form
extremal $SU(2)\times U(1)$ black holes the two non-degenerate strings 
$s_2$ and $s_3$ are joint at 
$r_{\rm deg}$ and form a straight line, whereas string $s_1$ degenerates to a
point.
Finally, for $\alpha >\alpha^{\rm max}_{II}$ there are 
no non-Abelian gauge fields  outside the degenerate horizon
and so, the strings (in the outside regions) are 
degenerated to points.
In contrast, in the inside region three non-degenerated strings persists.

\begin{figure}[h!]
\parbox{\textwidth}{
\centerline{
{\epsfysize=12.0cm \epsffile{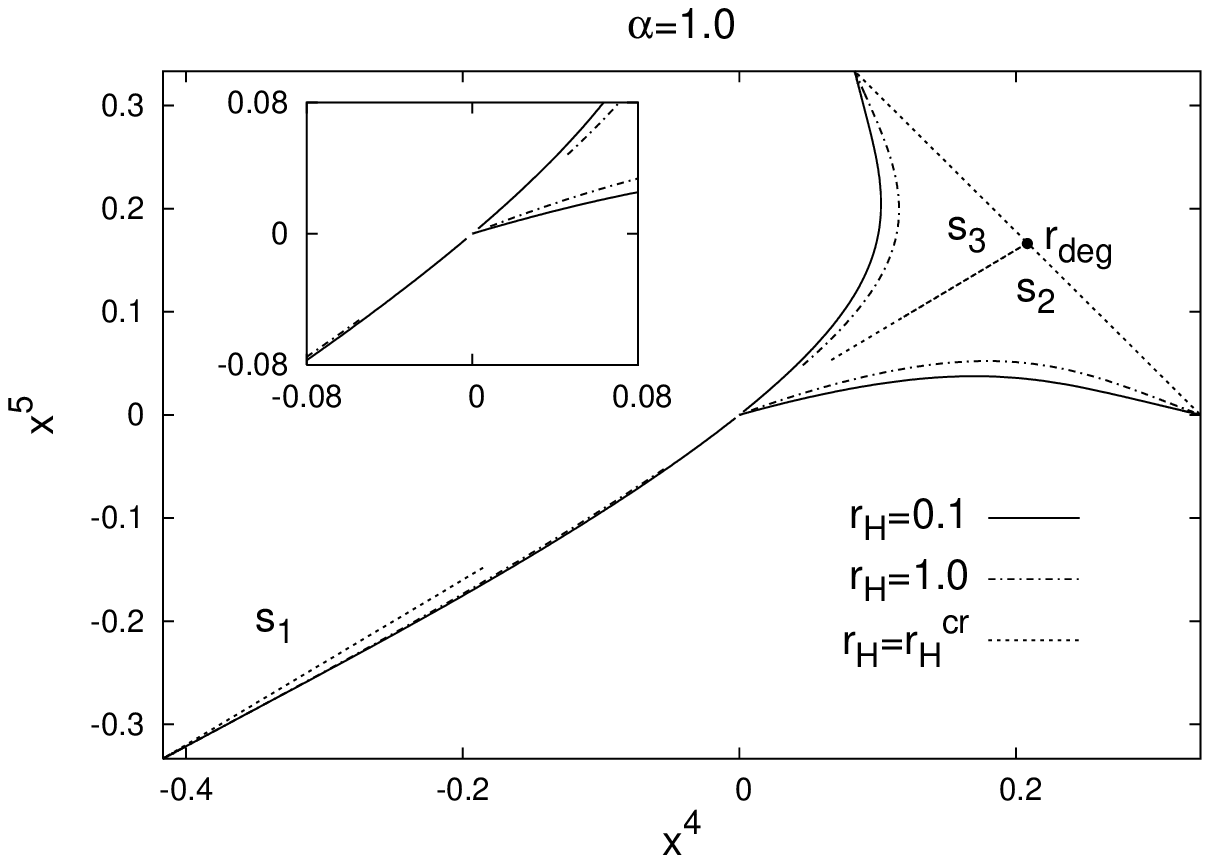} }
}\vspace{0.5cm}
{\bf Fig.~6} \small
Case III three strings  in the $x^4-x^5$ plane for 
$r_{\rm H}=0.1, 1.0, r_{\rm H}^{(cr)}$ 
and $\alpha=1$.
\vspace{0.5cm}
}
\end{figure}

\section{Extreme SU(4) Black Hole Solutions}

In section {\bf 3.2.2} it was shown that a special case 
occurs when one of the gauge functions becomes zero leading to  
 $SU(2)\times U(1)$ solutions with an additional magnetic charge.
Here we extend this work  in the $SU(4)$ case to construct  $SU(3)\times U(1)$
solutions.

The energy-momentum tensor, for purely magnetic gauge potential,  is
\bea
T^0_0-T^r_r & = &
\fr{4\kk_1B}{r^2}\left(3c_0'^2+4c_1'^2+3c_2'^2\right)\nonumber\\
&&+2\kk_2B
\sum_I\left(\fr{3}{4}(b^I_0)'^2+(b^I_1)'^2+\fr{3}{4}(b^I_2)'^2
           +(b^I_0)'(b^I_1)'+\fr{1}{2}(b^I_0)'(b^I_2)'+(b^I_1)'(b^I_2)'\right)
\nonumber\\
T^0_0 & = &
\fr{2\kk_1B}{r^2}\left(3c_0'^2+4c_1'^2+3c_2'^2\right)
+\fr{\kk_2}{r^2}\sum_I\left(
3c_0^2(b^I_0)^2+4c_1^2 (b^I_1)^2+3c_2^2 (b^I_2)^2\right)
\nonumber\\
& &
+\kk_2B\sum_I\left(
\fr{3}{4}(b^I_0)'^2+(b^I_1)'^2+\fr{3}{4}(b^I_2)'^2
           +(b^I_0)'(b^I_1)'+\fr{1}{2}(b^I_0)'(b^I_2)'+(b^I_1)'(b^I_2)'\right)
\nonumber\\
&&+\fr{2\kk_1}{r^4}
\left(\fr{9}{2}c_0^4+8c_1^4+\fr{9}{2}c_2^4-3c_0^2-4c_1^2-3c_2^2-6c_0^2c_1^2-
6c_2^2c_1^2+5\right), 
\eea
while the equations of motion for the matter profile functions are
\bea
\fr{1}{A}\left(ABr^2 \,{b^I_0}'\right)'&=&6c_0^2 b^I_0-4c_1^2 b^I_1
\nonumber\\
\fr{1}{A}\left(ABr^2\,{b^I_1}'\right)'&=&8c_1^2b^I_1-3c_0^2b^I_0-3c_2^2b^I_2
\nonumber\\
\fr{1}{A}\left(ABr^2\,{b^I_2}'\right)'&=&6c_2^2b^I_2-4c_1^2b^I_1
\nonumber\\
\fr{1}{A}\left(AB\,c_0'\right)'
&=&
c_0
\left[\fr{1}{r^2}(3c_0^2-2c_1^2-1)+\fr{\kk_2}{2\kk_1} \sum_I(b^I_0)^2\right]
\nonumber\\
\fr{1}{A}\left(AB\,c_1'\right)'
&=&
c_1
\left[\fr{1}{r^2}(4c_1^2-\frac{3}{2}c_0^2-\frac{3}{2}c_2^2-1)
+\fr{\kk_2}{2\kk_1}\sum_I(b^I_1)^2\right]
\nonumber\acc
\fr{1}{A}\left(AB\,c_2'\right)'
&=&c_2\left[\fr{1}{r^2}(3c_2^2-2c_1^2-1)
+\fr{\kk_2}{2\kk_1} \sum_I(b^I_2)^2\right].
\eea

Note that, setting $c_0(r)=0$, the Higgs field equations imply
\be
b^I_0(r) = \mbox{const}-\frac{2}{3} b^I_1(r) -\frac{1}{3} b^I_2(r),
\ee
and scaling the  gauge field functions as
\be
c_1(r)=w_0(r)/\sqrt{2}, \hs c_2(r)= w_1(r)\sqrt{2/3} \ , 
\ee
the $SU(4)$ equations transform to the $SU(3)$ ones  when 
$c_a, b^I_a$ are replaced by $w_a, h^I_a$, respectively.
However, there is the extra term $-6/(2r^4)$ in  
$T_0^0$  due to  the $U(1)$ field with charge $\sqrt{6}$.
Consequently, solutions do  exist only if $r_{\rm H} \geq \sqrt{6} \alpha$.

In the extreme case (ie. for $r_{\rm H} = \sqrt{6} \alpha$) the boundary 
conditions
for the gauge and  Higgs functions  at the horizon are
\be
w_0(r_{\rm H})=w_1(r_{\rm H})=1, \hs
h^I_0(r_{\rm H})=h^I_1(r_{\rm H})=0
\ee
while the string coordinates in transverse space are
\bea
(x^4_1(r) \ , \ x^5_1(r))  & =  & \vec{C} 
\nonumber \\
(x^4_2(r) \ , \ x^5_2(r))  
& =  & 
-\fr{1}{3}\,\vec{C} +\fr{1}{3} \left( 2 h^1_0(r) + h^1_1(r)\ , \ 
 2 h^1_0(r) + h^1_1(r) \right)
\nonumber \\
(x^4_3(r) \ , \ x^5_3(r))  
& =  & 
-\fr{1}{3}\,\vec{C} +\fr{1}{3} \left( -h^1_0(r) + h^1_1(r)\ , \  
-h^1_0(r) + h^1_1(r) \right) \nonumber \\
(x^4_4(r) \ , \ x^5_4(r))  
& =  & 
-\fr{1}{3}\,\vec{C} +\fr{1}{3} \left( -h^1_0(r) + 2 h^1_1(r)\ , \  
-h^1_0(r) + 2 h^1_1(r) \right).
\eea
Here ${\vec C}=\mbox{const}$ which implies 
that the first string degenerates to a point.

We computed the extreme solutions for a wide range of values of 
 $\alpha$, except the limit $\alpha \ra 0$.
In this region, only one branch has been found  which terminates at the
maximal value $\alpha_{\max}\approx 0.656$; while
at the limit $\alpha \ra \alpha_{\max}$  the formation   
of a (second) degenerate horizon  at $r_{\rm deg}=3 \alpha$ has been observed. 
In the outside region,   $r_{\rm deg} < r < \infty$, 
the limiting solution corresponds to an extremal $SU(2)\times U(1)$ black hole
and the $U(1)$ field has  charge $Q=3$ since  
$2 h^I_0(r)+ h^I_1(r)$ is constant; $w_0(r)$ vanishes identically;
 whereas $w_1(r)$ is non-trivial.
In the inside region, $r_{\rm H} \leq r < r_{\rm deg}$, the functions of 
the limiting solutions interpolate continuously between their values at 
$r_{\rm H}$ and $r_{\rm deg}$.
A degenerate horizon is formed since by increasing the coupling to gravity 
the gravitational radius becomes of the 
order of the core of the heaviest gauge field component. But the gauge
field components with smaller mass exist outside the degenerate horizon and 
that is why  the transition to an extremal $SU(2)\times U(1)$ black hole in 
the outside region occurs.

The formation of the degenerate horizon do affect the strings as shown in
Fig.~7 which presents the three strings in the $x^4-x^5$ plane 
for several values of $\alpha$ when  $\vec{C}=0$.
Note that,  for $\alpha$ small,
 the strings are similar to those of the globally regular solutions;
however as $\alpha$ increases the strings deform and 
in the limit  $\alpha\ra \alpha_{\max}$ the picture changes 
drastically.
For $r_{\rm deg} < r < \infty$ the second string consists of a single point 
$(x^4_2(\infty),x^5_2(\infty))$ which tends to the origin
as $r$ decreases from $r_{\rm deg}\ra r_{\rm H}$.
The third and fourth string form straight lines for $r_{\rm deg} < r < \infty$
and finally, merge when $r\ra r_{\rm deg}$. 
In the limit, $r \ra r_{\rm H}$ these two strings merge and form a 
straight line passing through origin.

\begin{figure}[h!]
\parbox{\textwidth}{
\centerline{
{\epsfysize=12.0cm \epsffile{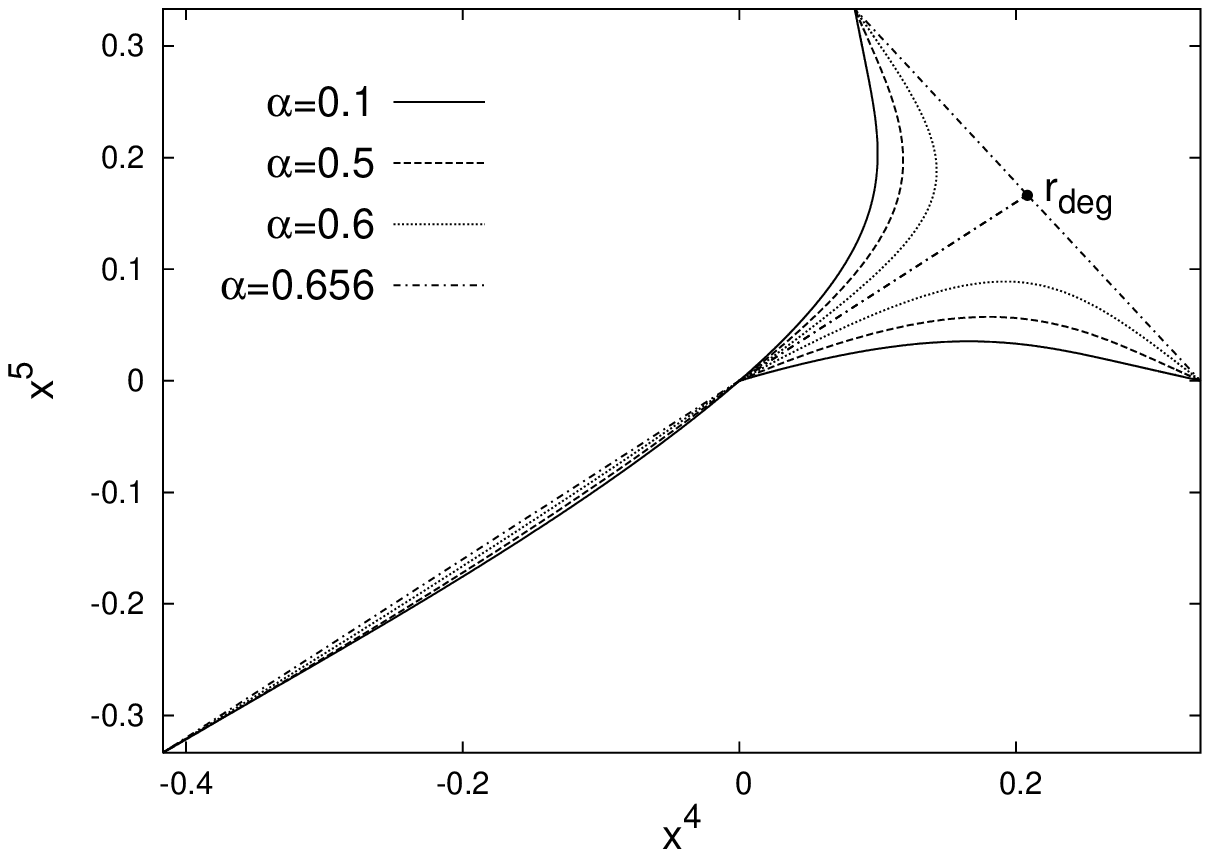} }
}\vspace{0.5cm}
{\bf Fig.~7} \small
The three strings of the extremal $SU(3)\times U(1)$ solutions 
in the $x^4-x^5$ plane for 
$\alpha=0.1, 0.5, 4, 0.6$ and $\alpha_{\rm max}=0.656$.
\vspace{0.5cm}}
\end{figure}

\section{Conclusion}

String theory assumes that space-time possess more than four dimensions
and that  these dimensions are  compactified on a scale of 
the Planck length.
The extra dimensions offer a solution to the hierarchy problem and 
assume that all known interactions (except gravity) are confined on a 
three-dimensional brane of a $(4+n)$-dimensional spacetime.
Therefore, it would be of interest to  construct our string and black hole
solutions from the corresponding Yang-Mills-Einstein equations
defined  in higher dimensions.
It has been shown in \cite{HY}, that an  $(4+n)$ dimensional 
 Yang-Mills-Einstein model  exists with 
$n$ Higgs fields  and $n$ dilatons \cite{HY} as long as the metric and 
 matter fields are  independent of the extra coordinates.
Thus, it would be interesting to construct our solutions by solving
the $(4+2)$-dimensional Yang-Mills-Einstein equations.

Also, an open and more difficult task would be to derive the same solutions
by solving the corresponding ${\cal N}=4$ supergravity equations of motion.\\

{\bf Acknowledgment}\\

The authors acknowledge valuable discussion with David Tong and 
Tassos Petkou.
BK gratefully acknowledges support by the DFG.

\end{document}